\documentclass[twocolumn]{aastex631}

\usepackage{mathtools}
\usepackage{amssymb}
\usepackage{appendix}
\shorttitle{AASTeX v6.3.1 Sample article}
\shortauthors{Qiao et al.}
\graphicspath{{./}{figures/}}

\begin{document}
\title{Chemo-dynamical Nature of the Anticenter Stream and Monoceros Ring\\}

\author{Yi Qiao}
\author{Baitian Tang}
\affiliation{School of Physics and Astronomy, Sun Yat-sen University, Daxue Road, Zhuhai, 519082, P.R.China; email:tangbt@sysu.edu.cn}
\affiliation{CSST Science Center for the Guangdong-Hong Kong-Macau Greater Bay Area, Zhuhai, 519082, P.R.China}

\author{Jianhui Lian}
\affiliation{South-Western Institute For Astronomy Research, Astronomy Building, Yunnan University, Chenggong District, Kunming, 650500, P.R.China}

\author{Jing Li}
\affiliation{School of Physics and Astronomy, China West Normal University, 1 ShiDa Road, Nanchong, 637002, P.R.China}

\author{Cheng Xu}
\affiliation{School of Physics and Astronomy, Sun Yat-sen University, Daxue Road, Zhuhai, 519082, P.R.China; email:tangbt@sysu.edu.cn}
\affiliation{CSST Science Center for the Guangdong-Hong Kong-Macau Greater Bay Area, Zhuhai, 519082, P.R.China}

\received{2023 September 6} 
\revised{2023 November 3}
\accepted{2023 November 27}
\submitjournal{The Astrophysical Journal}

\begin{abstract}

$\hspace*{0.3cm}$In the epoch of deep photometric surveys, a large number of substructures, e.g., over-densities, streams, were identified. With the help of astrometry and spectroscopy, the community revealed a complex picture of our Milky Way (MW) after investigating their origins. Off-plane substructures Anticenter Stream (ACS) and Monoceros Ring (MNC), once considered as dissolving dwarf galaxies, were later found to share similar kinematics and metallicity with the Galactic outer thin disk. In this work, we aim to chemically tag ACS and MNC with high-accuracy abundances from the APOGEE survey. By extrapolating chemical abundance trends in the outer thin disk region ($10 < R_{GC} < 18$ kpc, $0 < |Z_{GC}| < 3$ kpc), we found that ACS and MNC stars show consistent chemical abundances as the extrapolating values for 12 elements, including C, N, O, Mg, Al, Si, K, Ca, Cr, Mn, Co and Ni. The similar chemical patterns indicate that ACS and MNC have similar star formation history as the MW outer thin disk, meanwhile, we also excluded their dwarf galaxy association, as they are distinctive in multiple chemical spaces. The ages of ACS and MNC stars are consistent with the time of the first Sgr dSph passage, indicating their possible connection. 

\end{abstract}


\section{Introduction} \label{sec:1}

$\hspace*{0.2cm}$Our Galaxy has been continuously accreting materials from its surrounding environment since its beginning. The Galactic disk were formed after the Galactic halo, given that we can now trace the formation of different components of the Milky Way (MW) by their stars. The existence of Galactic disk indicates that there has been no major merger in the last few Gyrs. The most recent minor merger event that we know of is the accretion of Sagittarius (Sgr) dwarf galaxy. This event is suggested to be related to several observational features: Sgr tidal stellar streams \citep{2010ApJ...718.1128L}, phase spiral features \citep{2018Natur.561..360A,Xuyan2020}, disk undulation \citep{2015ApJ...801..105X} and recent starbursts in the Galactic disk \citep{2020NatAs...4..965R,Lian2020}. However, due to its remote distance, the detailed structure of disk undulation is still under debate. Monoceros Ring, the Anticenter Stream and Triangulum-Andromeda Stellar Cloud have been advocated to be part of the disk undulation. However, at their Galactocentric distances, they are mixed with stellar streams formed by disrupted dwarf galaxies or globular clusters. 

$\hspace*{0.2cm}$Monoceros Ring (MNC, also known as GASS), first discovered by \cite{2002ApJ...569..245N} as an overdensity, is an annular substructure lying in the direction of anticenter at low Galactic latitudes (b) and centered around 170$^\circ$ of Galactic longitude (l) according to \cite{2012ApJ...757..151L, 2021A&A...646A..99R}.  Stars in MNC are mostly located in Galactocentric distances of $\sim$18 kpc \citep{2003ApJ...588..824Y,2012ApJ...757..151L}, and possibly moving towards lower Galactic latitude  \citep{2012ApJ...757..151L}. 

$\hspace*{0.2cm}$ The Anticenter Stream (ACS) was first discovered as a long substructure in the anti-center direction which may consist of three similar but intersecting stellar streams \citep{2006ApJ...651L..29G}. It is located in a higher Galactic latitude compared to MNC, $160^\circ$ $<$ l $<$ $220^\circ$, $25^\circ$ $<$ b $<$ $35^\circ$ \citep{2012ApJ...757..151L}, with a distance of about 8.9 kpc to the Sun \citep{2006ApJ...651L..29G}. The spatial structure of ACS is almost perpendicular to the anti-Galactic center direction, and it is moving in the same direction as other disk stars \citep{2008ApJ...689L.117G}.

$\hspace*{0.2cm}$Both ACS and MNC were considered to be associated with tidal-disrupted dwarf galaxies since their discoveries, given their similar sizes and masses \citep[e.g.][]{2016ApJ...825..140M,2006ApJ...651L..29G}.
However, subsequent studies revealed that these two substructures are more likely originated from the MW disk, and they are also connected. According to the ACS and MNC high-purity giant star samples identified by \cite{2021A&A...646A..99R},  these two substructures intersect the Galactic plane with small angles. The member stars of ACS and MNC show similar longitudinal proper motion distribution \citep{2020MNRAS.492L..61L}, and their kinematics are compatible with the Galactic rotation curve at those distances or slightly slower \citep{2021A&A...646A..99R}. 
Compared to kinematics, chemical features are more stable, and chemical abundances are more reliable to reveal the origin of a stellar system. The metallicities of ACS and MNC are suggested to be between -0.5 and -1.0 dex, with ages mostly between 6-10 Gyr, and [Mg/Fe] between 0 and 0.15 dex \citep{2020MNRAS.492L..61L,2021ApJ...910...46L}. Thus, these substructures showed significant resemblances with the MW disk. 

$\hspace*{0.2cm}$Based on the aforementioned literature, the star formation history of ACS and MNC still needs further investigation. Since star formation history will affect not only metallicity, but also other chemical abundances, this work aims to explore the detailed chemo-dynamical features of these two substructures, including multiple elements of different species. In \textbf{Section \ref{sec:2}}, we describe how to select ACS and MNC candidates and comparison sample from the APOGEE survey. In \textbf{Section \ref{sec:3}}, we carefully analyze the orbits and chemical abundances of the candidate stars. Their origins are discussed by comparing with the MW disk and nearby dwarf galaxies. The chemo-dynamical signatures and formation scenarios are discussed in \textbf{Section \ref{sec:4}}, and a brief summary is given in \textbf{Section \ref{sec:5}}. 

\section{Data}
\label{sec:2}

$\hspace*{0.2cm}$In this work, we first cross-matched the ACS and MNC candidates selected by \cite{2021yCat..36460099R} with APOGEE DR17 \citep{2015AJ....150..173N,2017AJ....154...94M}. The Ramos2021 samples with regular proper motions have also been used to provide stream ``tracks'' by \cite{2023MNRAS.520.5225M}. The stellar parameters and chemical abundances used in this work were extracted by the APOGEE Stellar Parameter and Chemical Abundance Pipeline \citep[ASPCAP,][]{2016AJ....151..144G}.  To insure the reliability of stellar parameters and chemical abundances, we implemented the following criteria:
\begin{itemize}
	\item ASPCAPFLAG = 0
	\item \textit{$v_{scatter}$ $<$ 1 km $s^{-1}$}
	\item \textit{$v_{err}$ $<$ 0.2 km $s^{-1}$}
	\item \textit{SNR $>$ 80}
	\item 3700 K $< T_{eff} <$ 5500 K
	\item parallax  $<$  0.333 mas
\end{itemize}

\begin{figure}[b]
	\raggedright
	\includegraphics[scale=0.57]{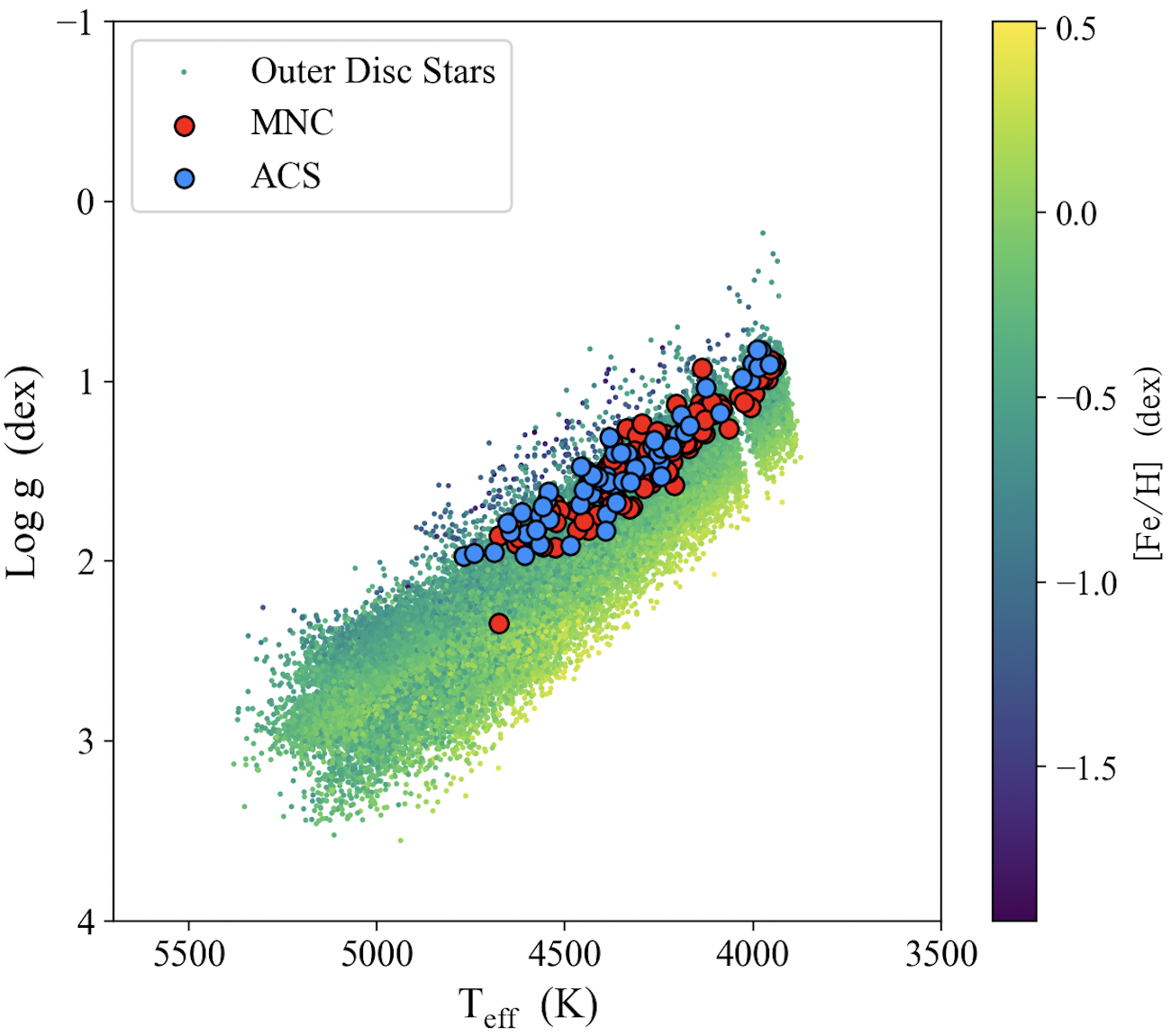}
	\caption{The spectroscopic Hertzsprung-Russell diagram of Log g \textit{vs} $T_{eff}$ obtained from ASPCAP for the outer disk (colored dots with a two-dimensional histogram where densely populated), ACS (blue circles), and MNC (red circles) samples. }
	\label{fig1}
\end{figure}

$\hspace*{0.2cm}$We further refined our sample using the spectrophotometric heliocentric distances from StarHorse \citep{starhorse2018}. Based on the distances of two substructures and their associated uncertainties, we only selected stars with Galactocentric radii ($R_{GC}$) larger than 14 kpc and corresponding errors smaller than 2 kpc. Finally, there are 59 ACS candidates and 116 MNC candidates left. 

\begin{figure*}[htbp]
	\centering
	\includegraphics[scale = 0.59]{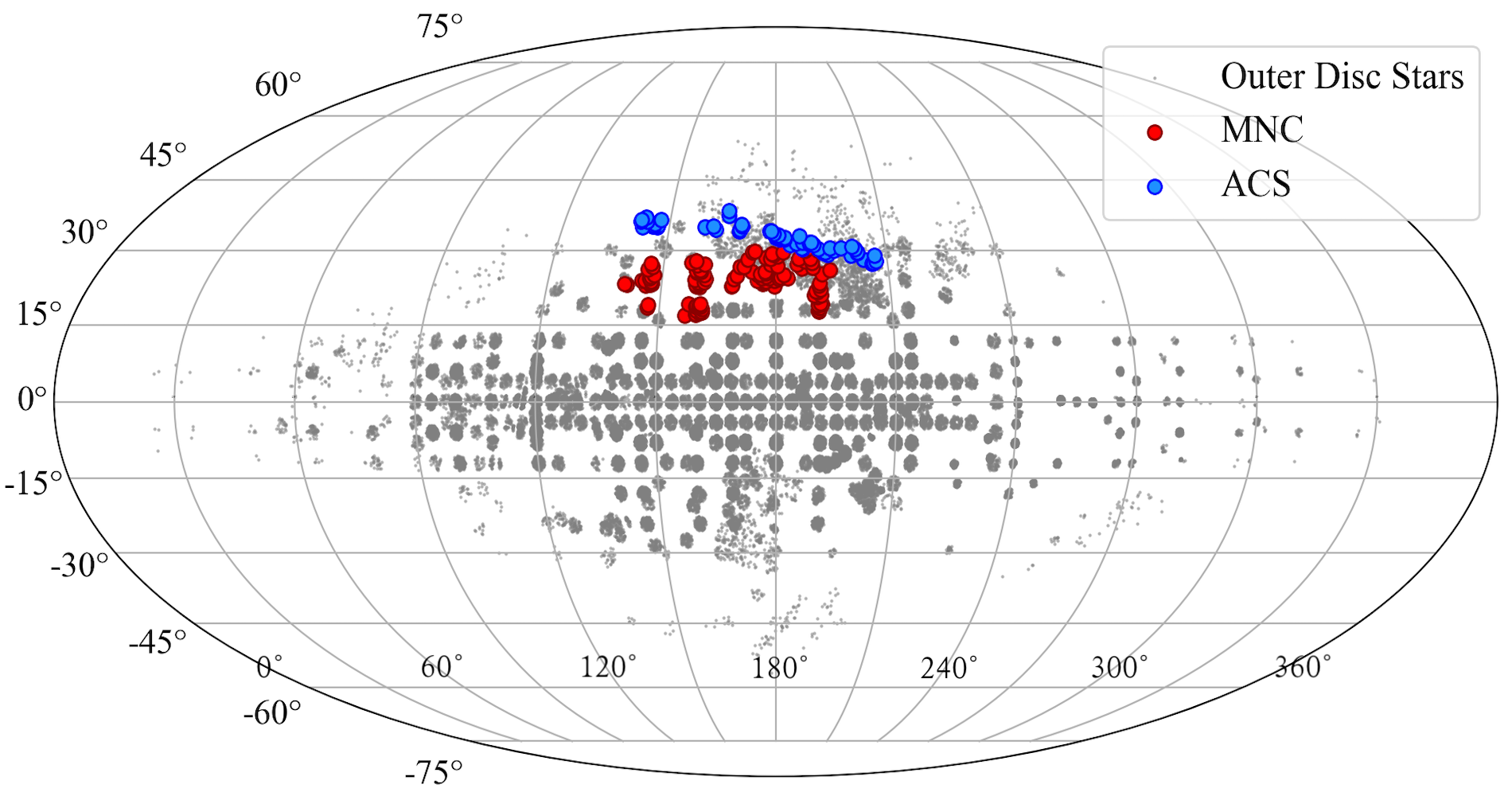}
	\caption{The Mollweide projection of our samples in the Galactic coordinates. Circles filled with blue and red stand for ACS and MNC stars respectively, and grey dots in the background represent the outer disc stars. }
	\label{fig2}
\end{figure*}

$\hspace*{0.2cm}$The ASPCAP first derives stellar atmospheric parameters by finding the best model fit to the entire APOGEE spectrum. Abundance of an individual element is derived from matching the observed feature lines with atmospheric models of different chemical abundances in spectral ``windows'' (a ``window'' means a wavelength range which is dominated by spectral features associated with a specific element). For some elements with low SNR spectral features, their abundances are less reliable. To avoid such situation, we checked the spectral features in all the windows of our sample stars. Several elements were removed due to low SNR absorption lines. In the end, we selected 12 elements with reliable abundances: C, N, O, Mg, Al, Si, K, Ca, Cr, Mn, Co, Ni. 

$\hspace*{0.2cm}$To investigate if ACS and MNC stars are chemically similar to MW disk/dwarf galaxies, we selected member stars of the MW outer disk and nearby dwarf galaxies for comparison. \textbf{a)} dwarf galaxies. Based on the member stars of Sgr dSph and GSE identified by \cite{2021ApJ...923..172H}, we implemented the aforementioned criteria to minimize their possible impact when discussing chemical abundances. A few hundred stars were selected as our dwarf galaxy sample. \textbf{b)} MW outer disk. Besides the aforementioned criteria, we further select stars with \textit{$R_{GC}$ $>$ 10 kpc}, \textit{$|Z_{GC}|$ $<$ 3 kpc} and \textit{$R_{err}$ $<$ 1 kpc, $Z_{err}$ $<$ 0.5 kpc}. More than fifty thousand stars were selected. As shown in the left panel of \textbf{Figure \ref{fig1}}, our outer disk sample covers the $T_{eff}$ and Log g range of the ACS and MNC samples, which makes the chemical abundance comparison more reliable. The distributions in Galactic coordinates of ACS and MNC samples are exhibited in \textbf{Figure \ref{fig2}}, and the outer disk stars are plotted as a contrast. 

\section{Results and Analysis}
\label{sec:3}

\subsection{Orbital Parameters} 

\begin{figure*}[htbp]
	\centering
	\includegraphics[scale=0.425]{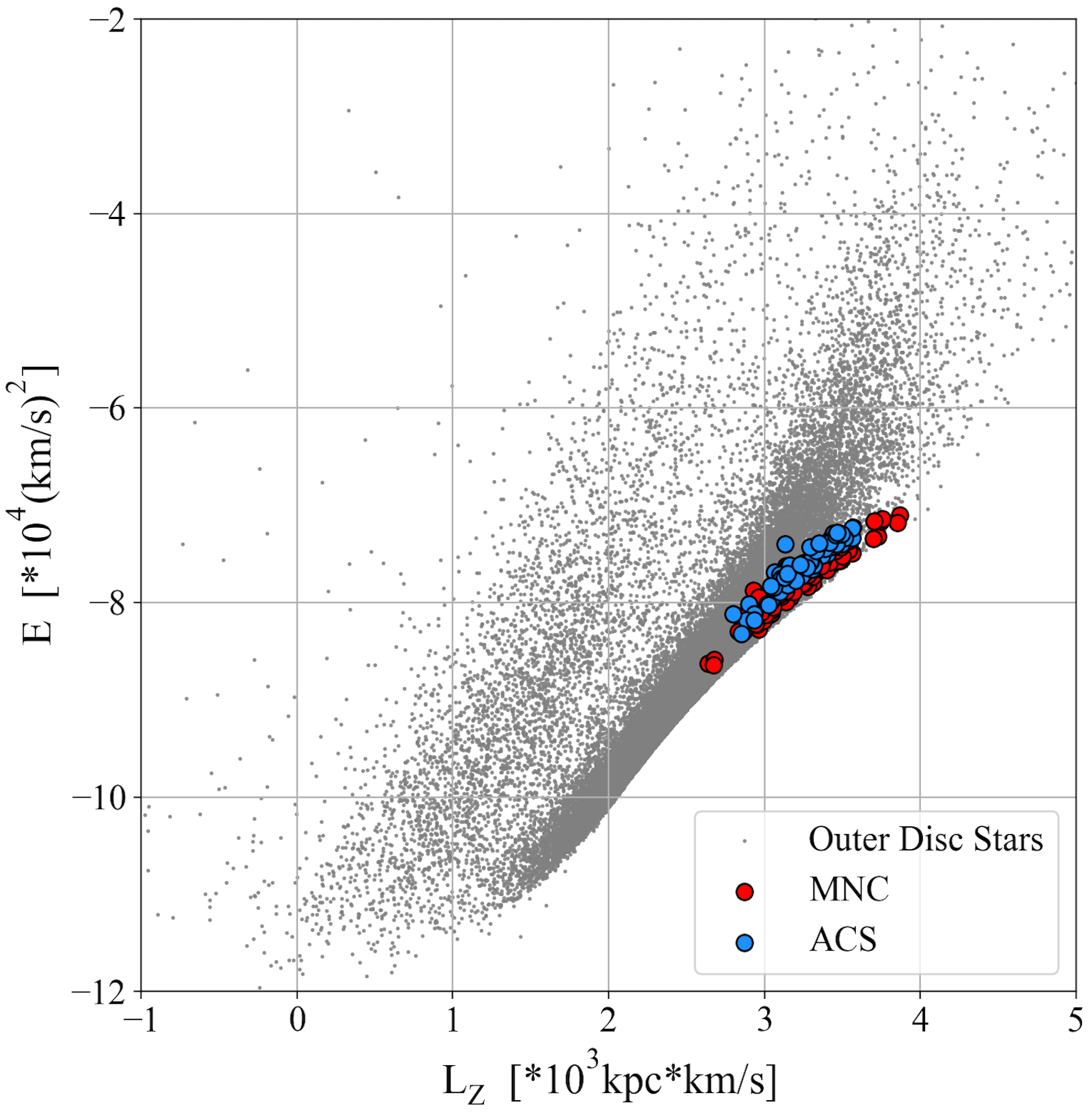}
	\hspace{0.7cm}
	\includegraphics[scale=0.425]{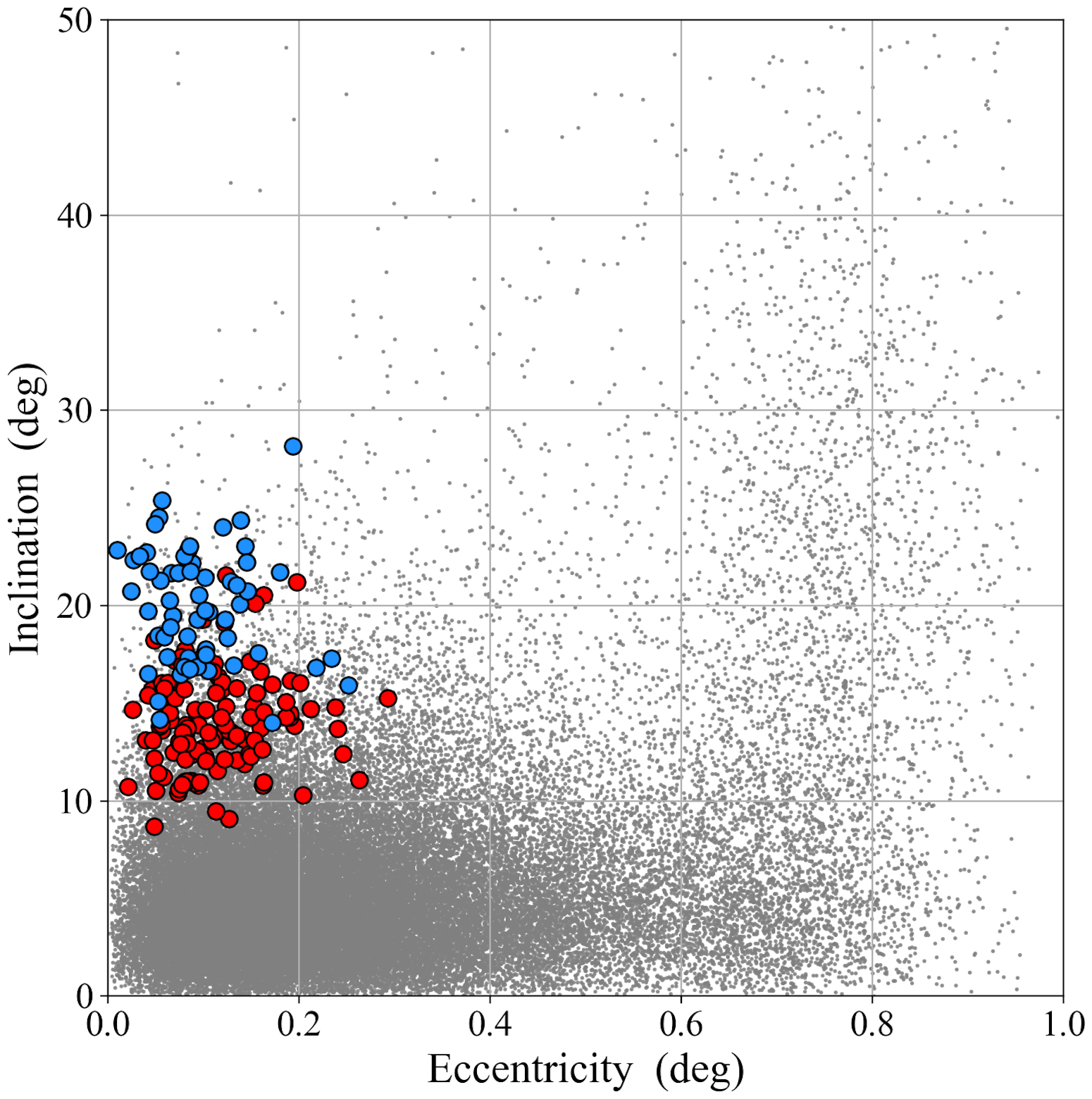}
	\caption{Left panel: The diagram of E \textit{vs} $L_Z$ of ACS, MNC and outer disk stars. MNC stars (red circles) and ACS stars (blue circles) form an approximate linear relationship, and they stay together with the majority of outer disk stars (grey dots) generally. Right panel: Corresponding orbital inclination and eccentricity of the three samples. }
	\label{fig3}
\end{figure*}

\begin{figure*}[htbp]
	\centering
	\includegraphics[scale=0.43]{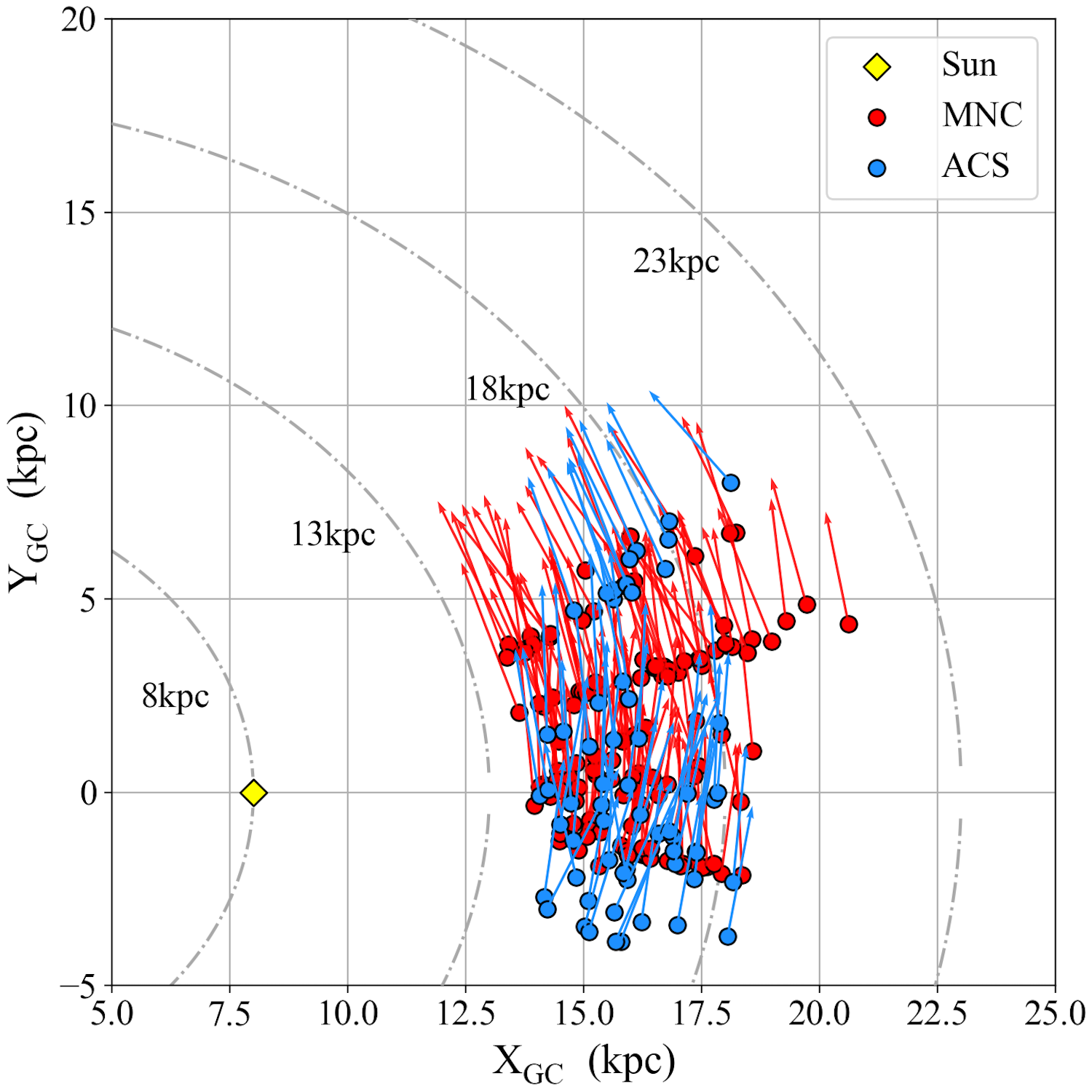}
	\hspace{0.68cm}
	\includegraphics[scale=0.43]{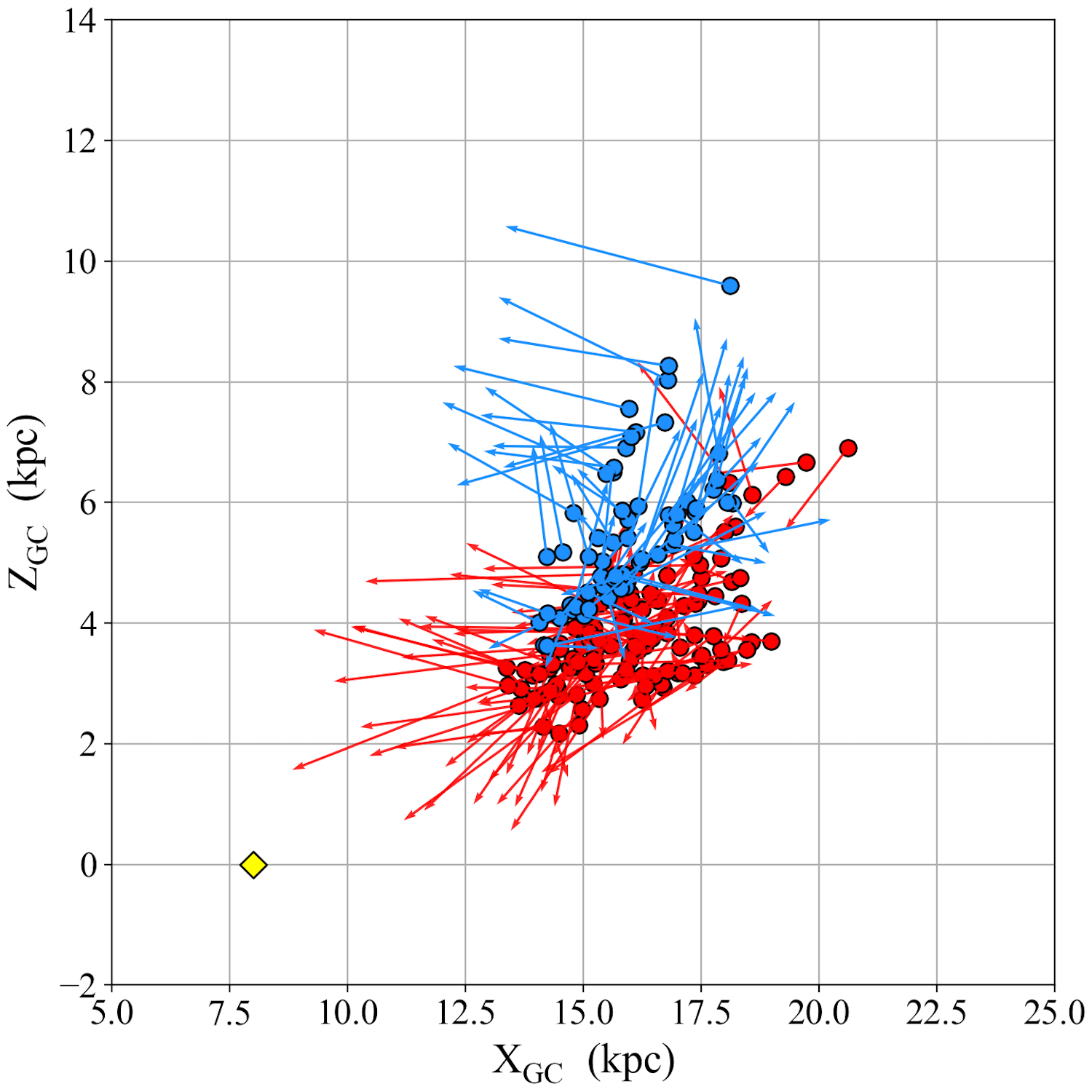}
	\caption{($X_{GC}$,$Y_{GC}$) (left panel) and ($X_{GC}$,$Z_{GC}$) (right panel) projections of ACS and MNC stars, with vectors representing their velocities in the X, Y and Z Galactocentric Cartesian directions. The orbital parameters of ACS stars are represented in blue and MNC stars are in red. }
	\label{figa4}
\end{figure*}

$\hspace*{0.2cm}$To confirm stars in each of the ACS and MNC samples are dynamically similar to each other, we calculated their orbital parameters with the publicly available Python library, GALPY \citep{2015ApJS..216...29B}. We adopted a steady-state and axisymmetric MW potential: MWPotential2014. The values of the solar Galactocentric radius and the local circular velocity are \textit{$R_\odot$ = 8 kpc}, \textit{$v_\odot$ = 220 km $s^{-1}$} respectively. And the peculiar velocity of the Sun is [11.1, 12.24, 7.25] km $s^{-1}$ \citep{2010MNRAS.403.1829S}. 

\begin{figure*}[htbp]
	\centering
	\includegraphics[scale=0.7]{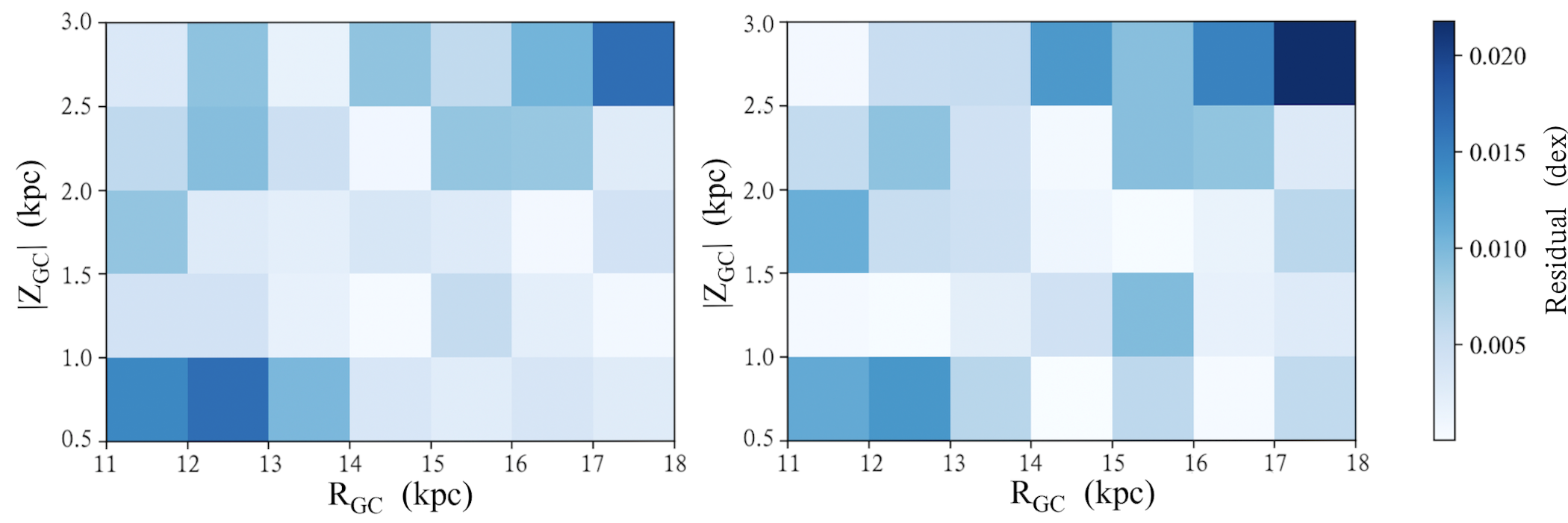}
	\caption{The residuals of [Mg/Fe] in different fitting models at different $R_{GC}$ and $Z_{GC}$, the left panel shows results of our model: [X/Y] = a$\times\sqrt{R_{GC}}$ + b $\times \sqrt[4]{|Z_{GC}|}$ + c , and the right panel exhibits those of the linear model: [X/Y] = a$\times$$R_{GC}$ + b$\times$$|Z_{GC}|$ + c . }
	\label{fignew}
\end{figure*}

$\hspace*{0.2cm}$The orbital parameters of the stars studied in this work as well as their spatial projections and velocity vectors are shown in \textbf{Figure \ref{fig3}} and \textbf{Figure \ref{figa4}}. In general, most ACS and MNC stars are dynamically consistent with the outer disk in total orbital energy (E) \textit{vs} vertical angular momentum ($L_Z$) space. The orbital inclinations of ACS stars ($\sim$ $21^{\circ}$) are systematically larger than those of MNC stars ($\sim$ $14^{\circ}$), which agrees with the fact that they are further away from the Galactic midplane (larger $Z_{GC}$ values). \cite{2016ApJ...823....4D} verified that passages of satellite galaxies like Sgr could induce tides that excite the orbits of the MW's in-situ population. They achieved this through the utilization of high-resolution N-body simulations. In this scenario, the non-zero orbital inclinations of stars in the ACS and MNC might arose as a consequence of interactions between Sgr and the MW. The orbital eccentricities of stars in both substructures are relatively small ($\sim$ 0.1), indicating near-circular orbits (also see their velocity vectors). One star from our ACS sample, with a larger value of inclination compared to other candidates, also exhibits a slightly higher E in the E \textit{vs} $L_Z$ diagram than the average distribution. However, this star is not clearly separable in the spatial projection of ACS and MNC stars, and it shows no significantly different chemical abundances compared to others, so we decided to keep it as one of the ACS members. 

\subsection{Chemical Abundances}

\begin{figure*}[t]
	\centering
	\includegraphics[scale = 0.458]{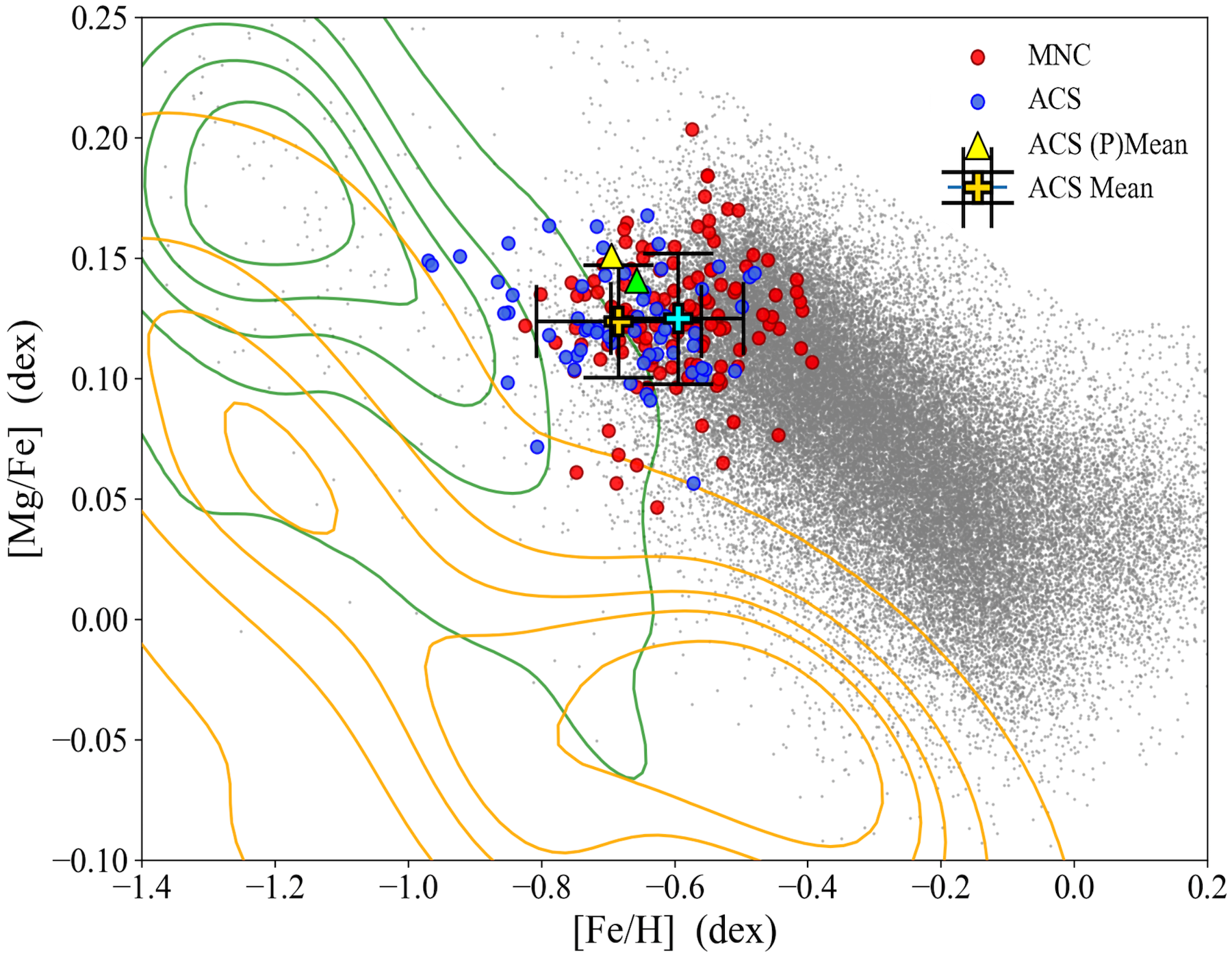}
	\hspace{0.34cm}
	\includegraphics[scale=0.458]{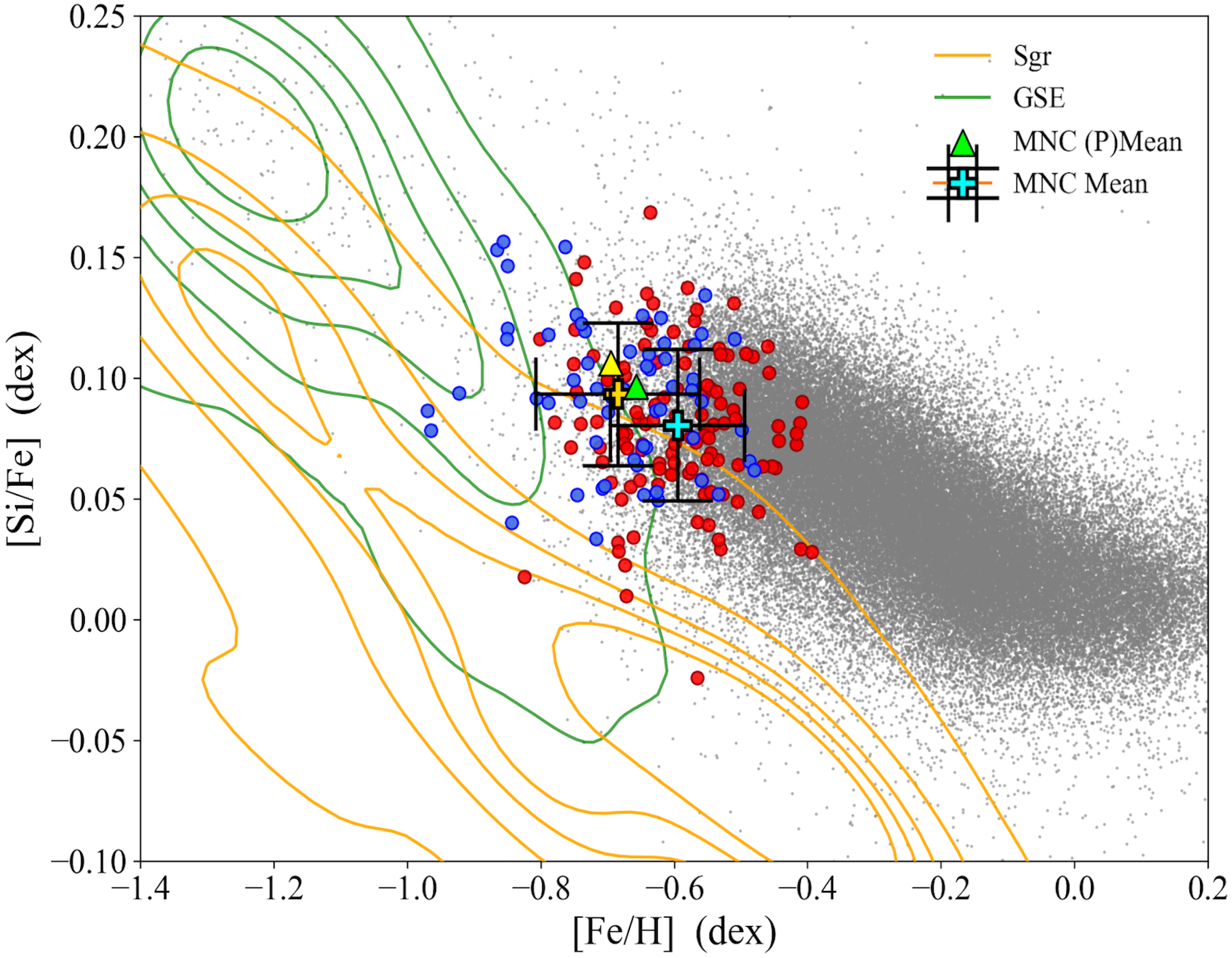}
	\includegraphics[scale=0.458]{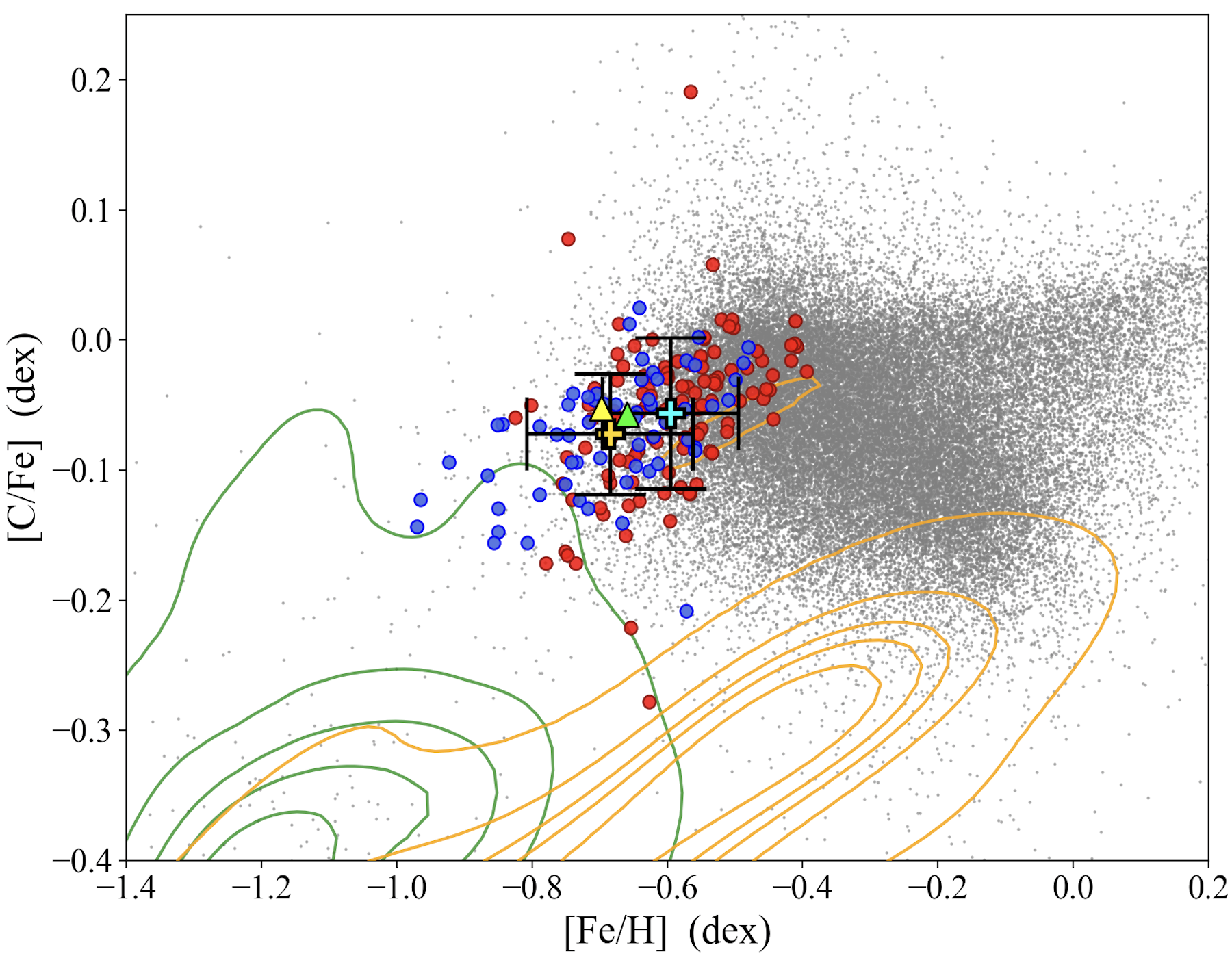}
	\hspace{0.365cm}
	\includegraphics[scale = 0.458]{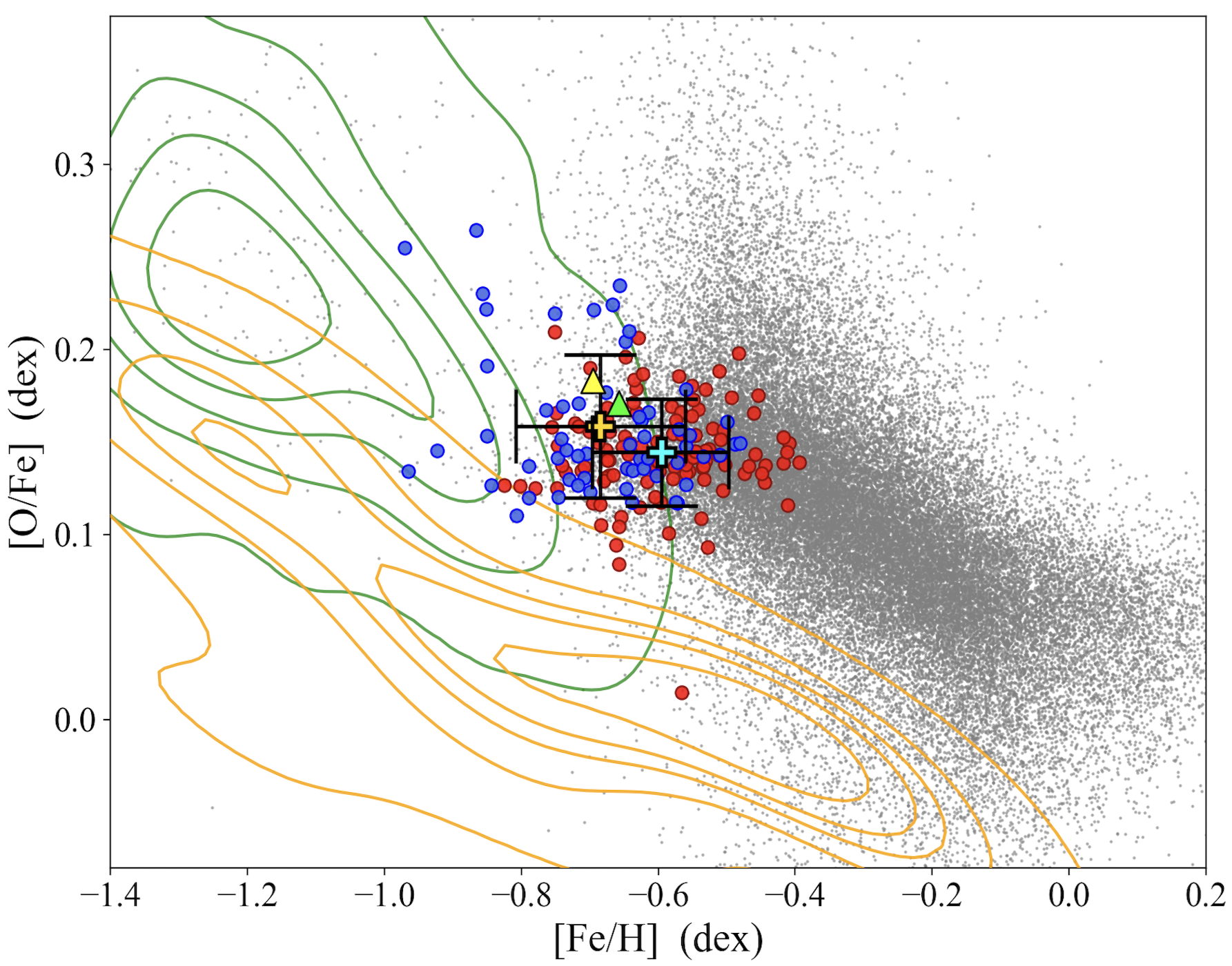}
	\includegraphics[scale=0.458]{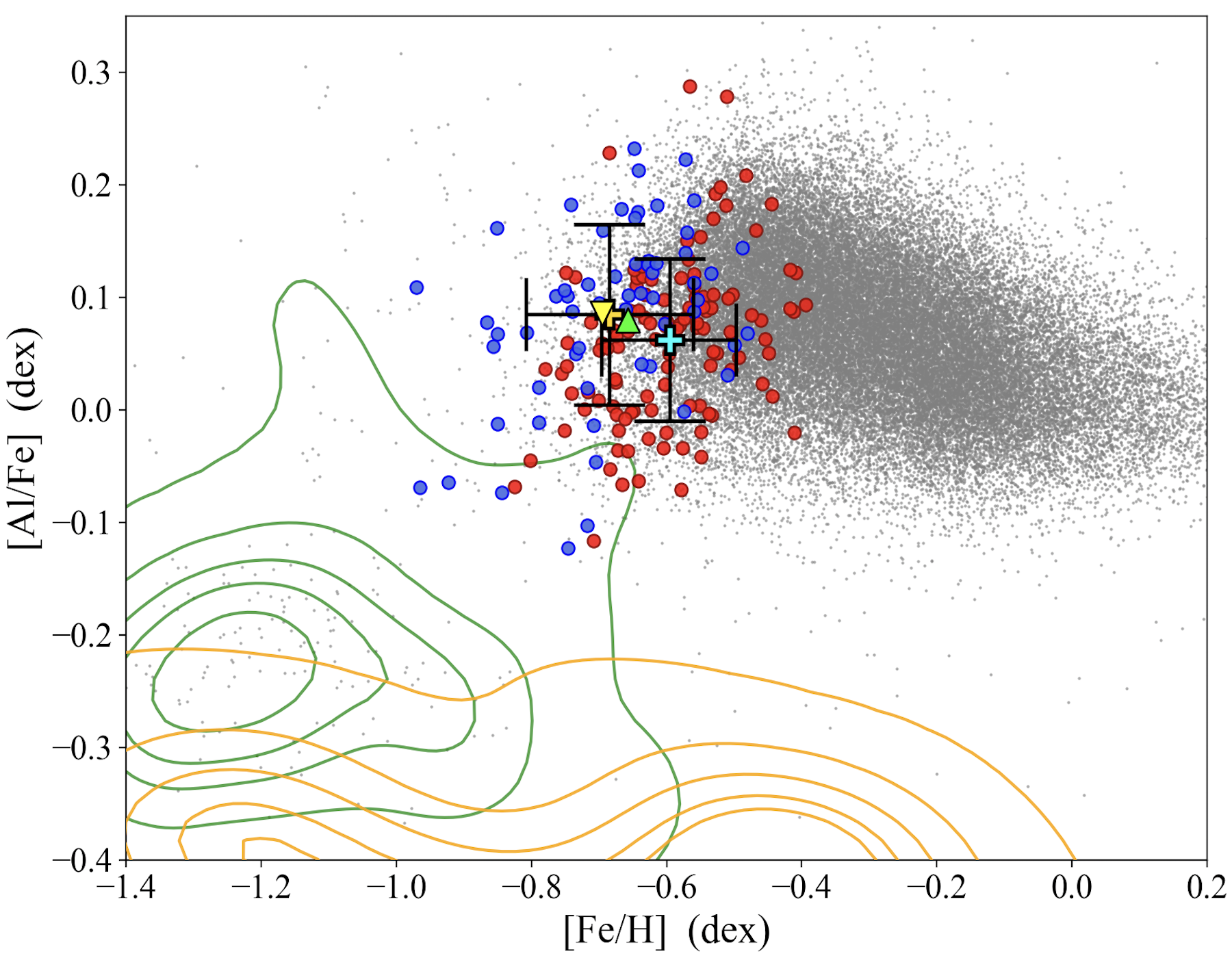}
	\hspace{0.34cm}
	\includegraphics[scale=0.458]{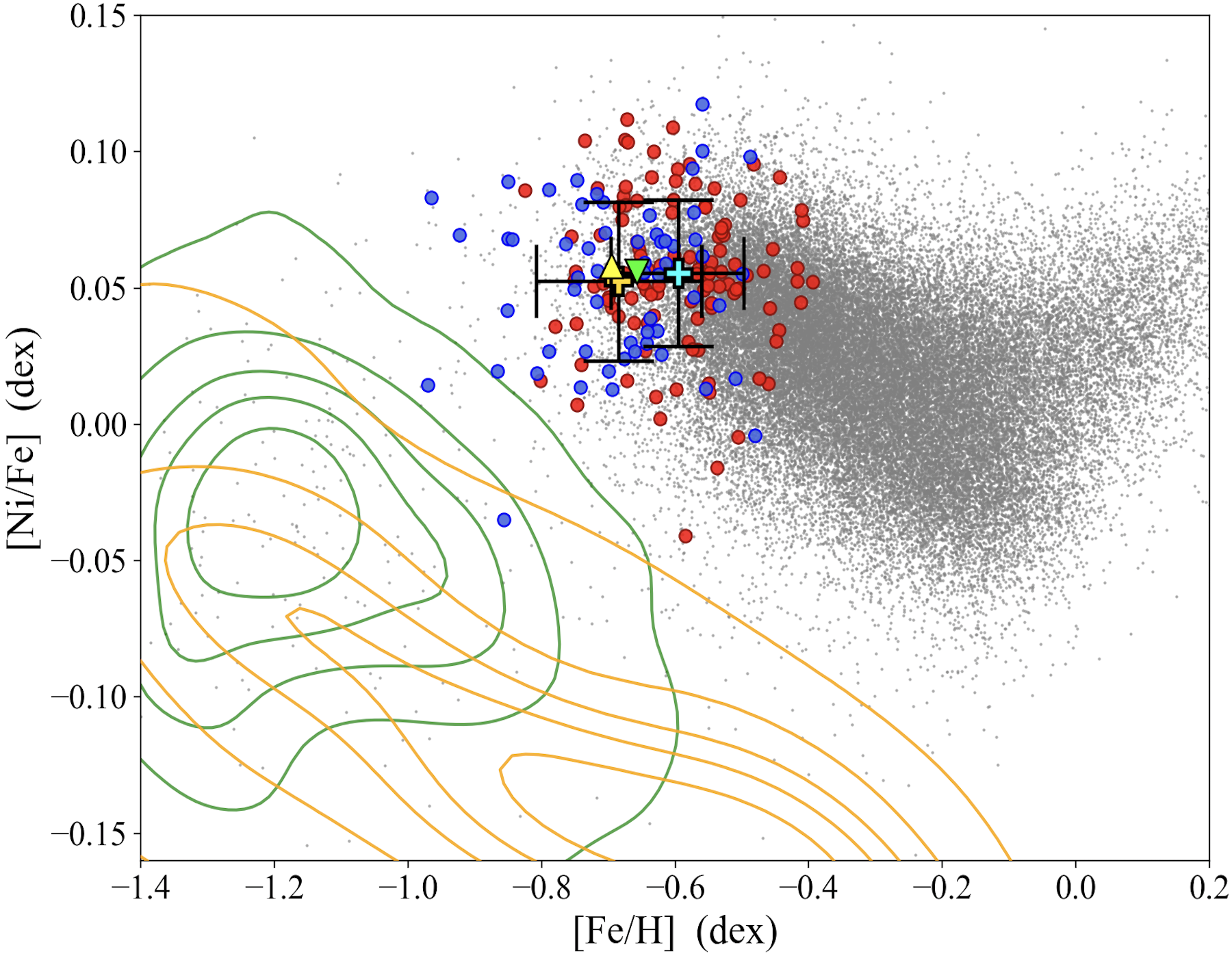}
	\caption{Abundance extrapolation results of 6 representative elements. Outer disk stars are plotted in grey and isodensity contours stand for Sgr (orange) and GSE (green) stars. ACS and MNC stars are represented by blue and red circles respectively. For each element, the crosses and errorbars' lengths show mean abundances and corresponding standard deviations of ACS (gold cross) and MNC (cyan cross) samples, meanwhile, two triangles represent their predicted mean abundances (yellow for ACS, lime for MNC). }
	\label{fig5}
\end{figure*}

$\hspace*{0.2cm}$If ACS and MNC are part of the Galactic outer disk, then chemical abundances of these substructures should follow the trend outlined by other Galactic outer disk stars. Inspired by the work of \cite{2018ApJ...859L...8H}, we fit the abundance ratios [X/Y] as a function of their Galactic locations. But instead of only considering $R_{GC}$, here we use two independent variables, $R_{GC}$ and $Z_{GC}$ for the following reasons: 1. Studies \citep[e.g.,][]{2022ApJ...928...23E} show that abundance variations in the MW disk depend on both $R_{GC}$ and $Z_{GC}$; 2. Using $R_{GC}$ as the only variable would predict almost same abundance ratios for both ACS and MNC stars, given their similar $R_{GC}$. But this clearly betrays our observation: ACS stars on average are more metal-poor ([Fe/H]) than MNC stars (Figure \ref{fig5}). Moreover, since most of the MW outer disk stars are thin disk stars, to avoid contamination from thick disk stars when fitting mathematical functions, we further exclude thick disk stars with large [Mg/Fe] at a given [Fe/H]. 

\begin{table*}[htbp]
	\raggedright
	\caption{Fitting coefficients of 13 elements' chemical trends in our model: [X/Y] = $a \times \sqrt{R_{GC}} + b \times \sqrt[4]{|Z_{GC}|} + c$ . }
	\label{tab:1}
	\begin{tabular}{cccccccccccccc}
		\hline\hline\noalign{\smallskip}	
		Coefficient & [Fe/H] & [Mg/Fe] & [Si/Fe]& [Ca/Fe] & [C/Fe] & [N/Fe] & [O/Fe] & [Al/Fe] & [K/Fe] & [Cr/Fe] & [Mn/Fe] & [Co/Fe] & [NiFe] \\
		\noalign{\smallskip}\hline\noalign{\smallskip}
		a ($\times10^{-2}$) & -23.88 & 1.474 & 0.081 & -2.076 & -2.939 & 8.067 & -0.108 & -2.682 & -1.846 & 0.679 & 2.920 & 8.457 & 3.854 \\
		b ($\times10^{-2}$) & -27.72 & 7.405 & 7.038 & 1.913 & 3.301 & -4.225 & 8.873 & 4.243 & 6.307 & -0.826 & -7.032 & 2.092 & 1.529 \\
		c ($\times10^{-2}$) & 69.12 & -2.128 & -0.428 & 8.849 & 1.559 & -1.671 & 5.214 & 12.90 & 9.809 & -4.608 & -6.804 & -30.17 & -12.06 \\
		\noalign{\smallskip}\hline
	\end{tabular}
\end{table*}

\begin{figure*}[htbp]
	\centering
	\includegraphics[scale=0.458]{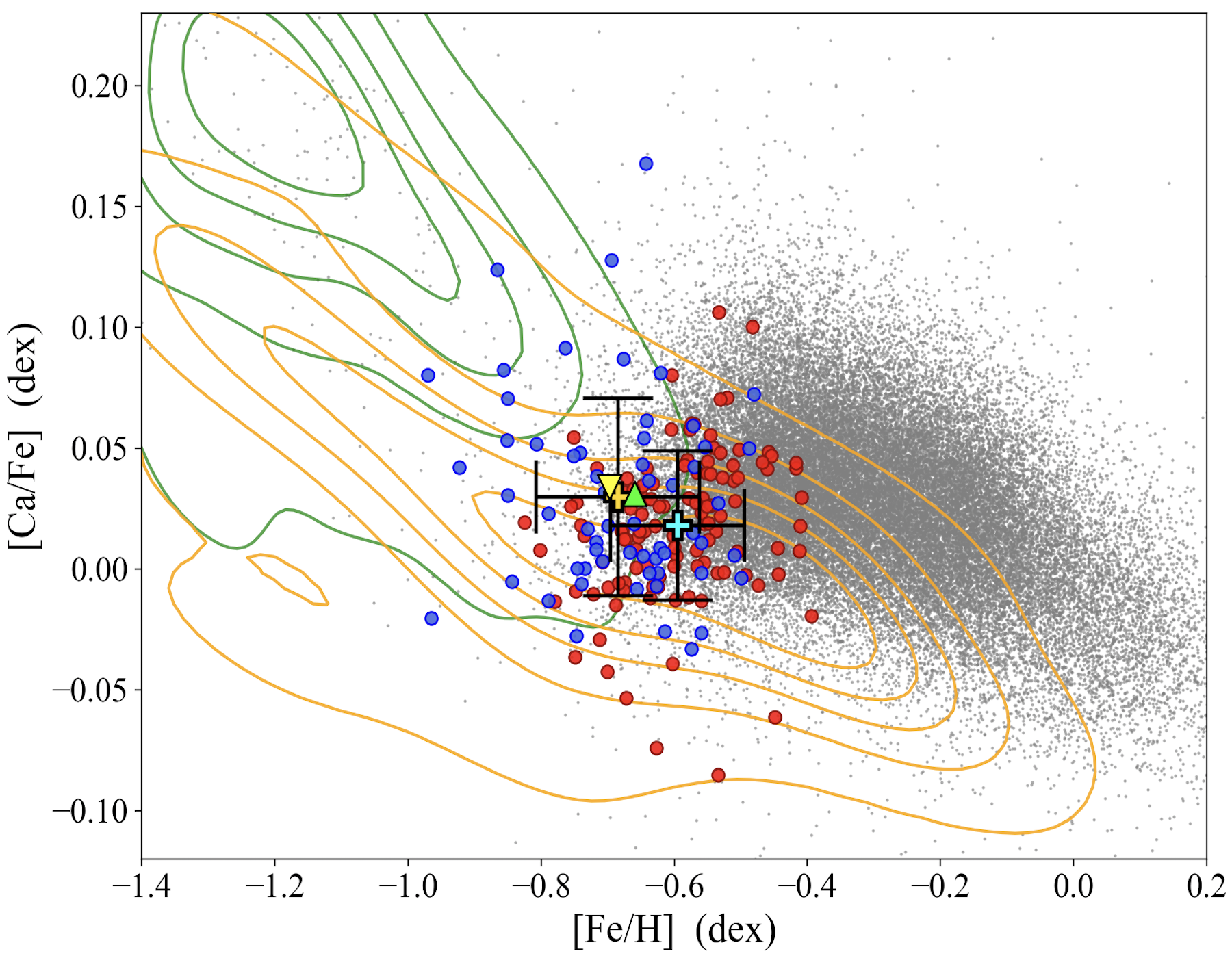}
	\hspace{0.34cm}
	\includegraphics[scale=0.458]{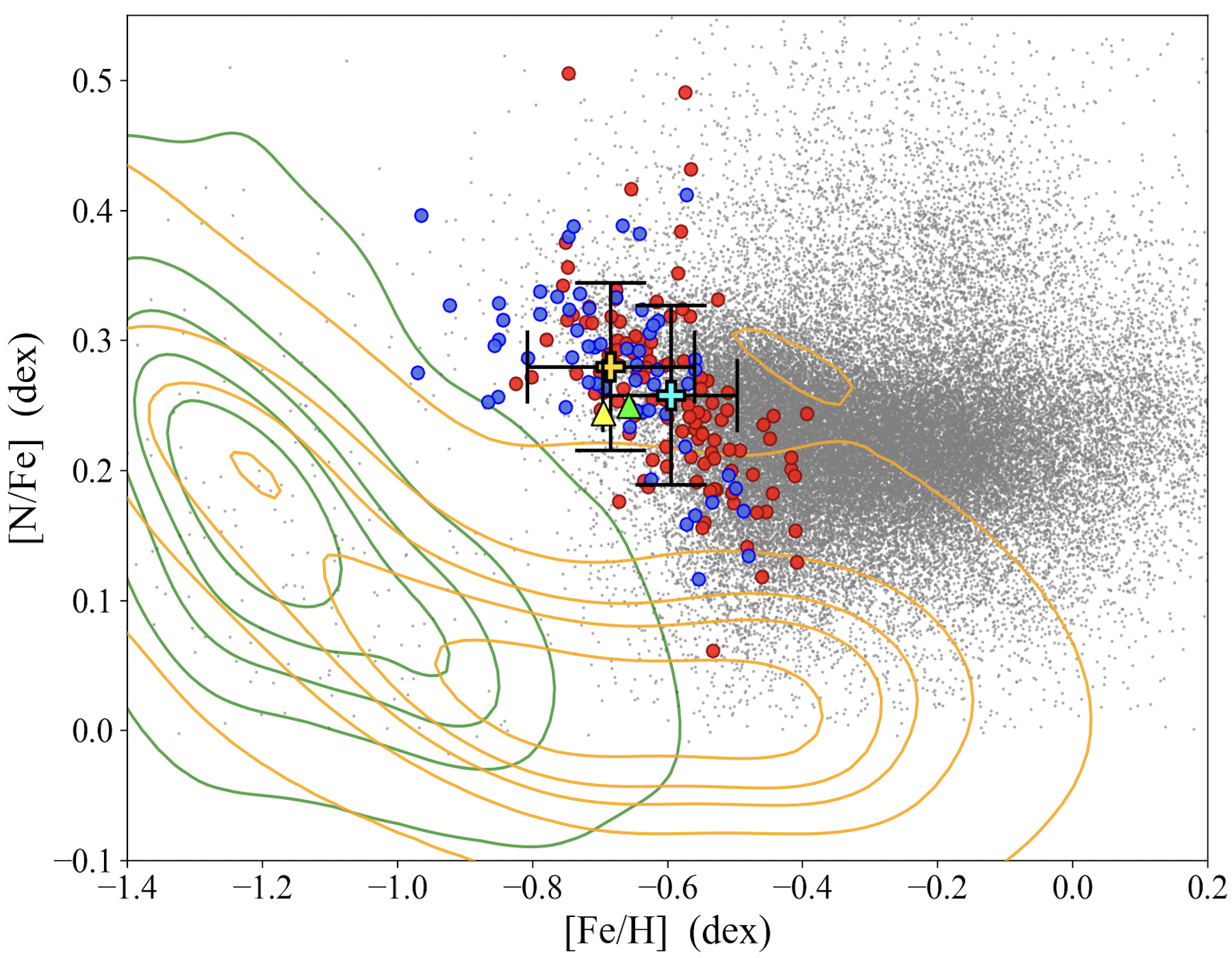}
	\includegraphics[scale=0.458]{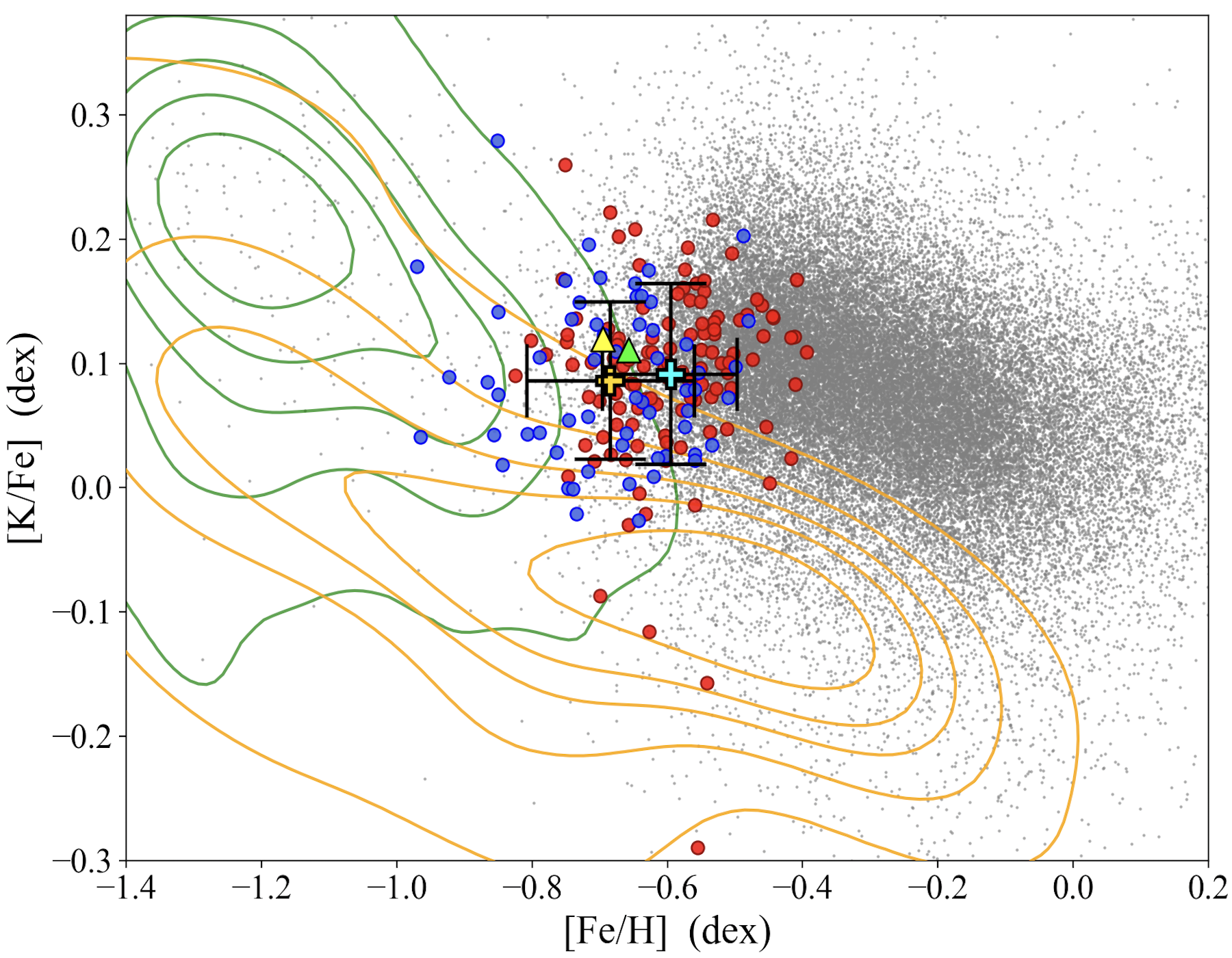}
	\hspace{0.34cm}
	\includegraphics[scale=0.458]{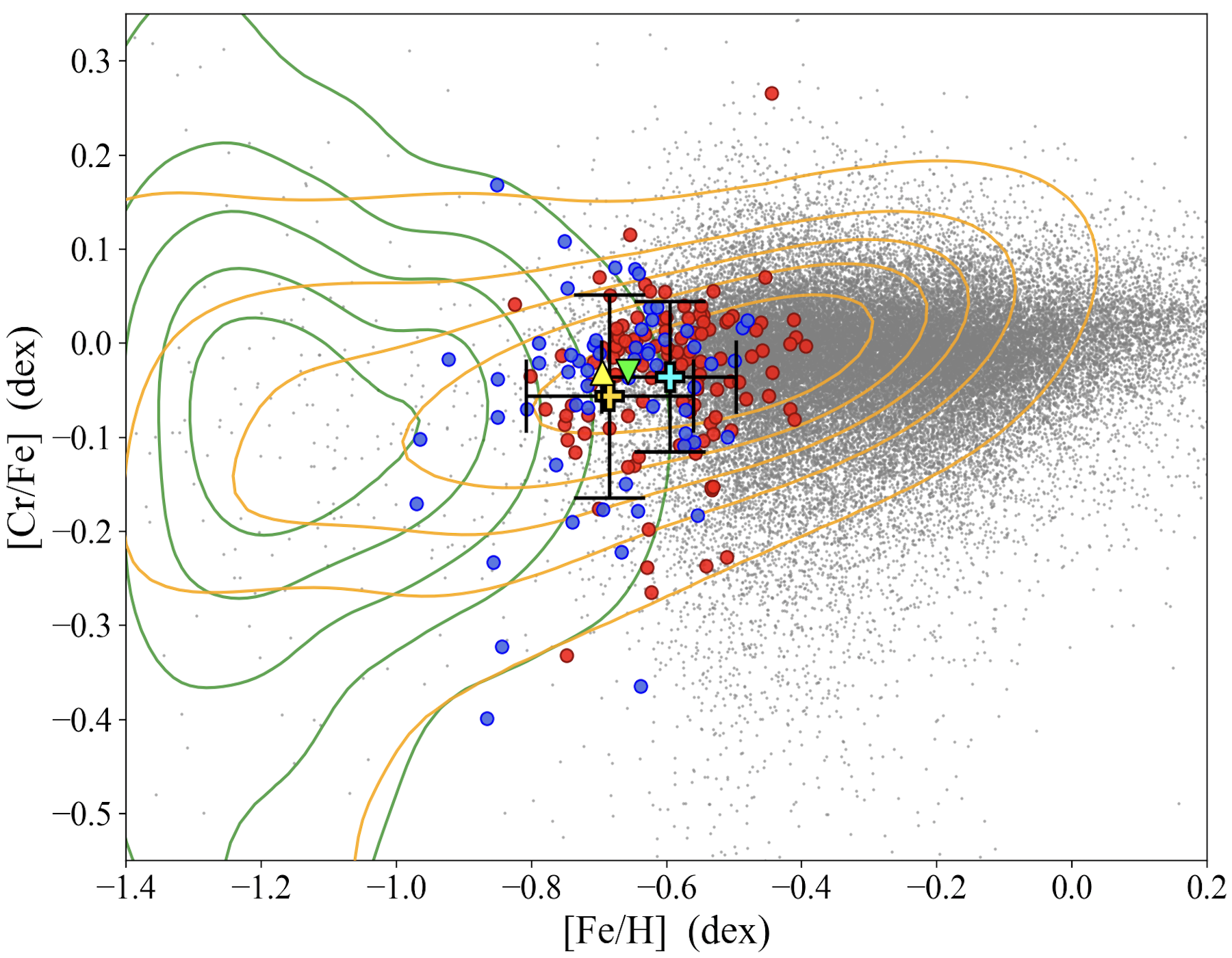}
	\includegraphics[scale=0.458]{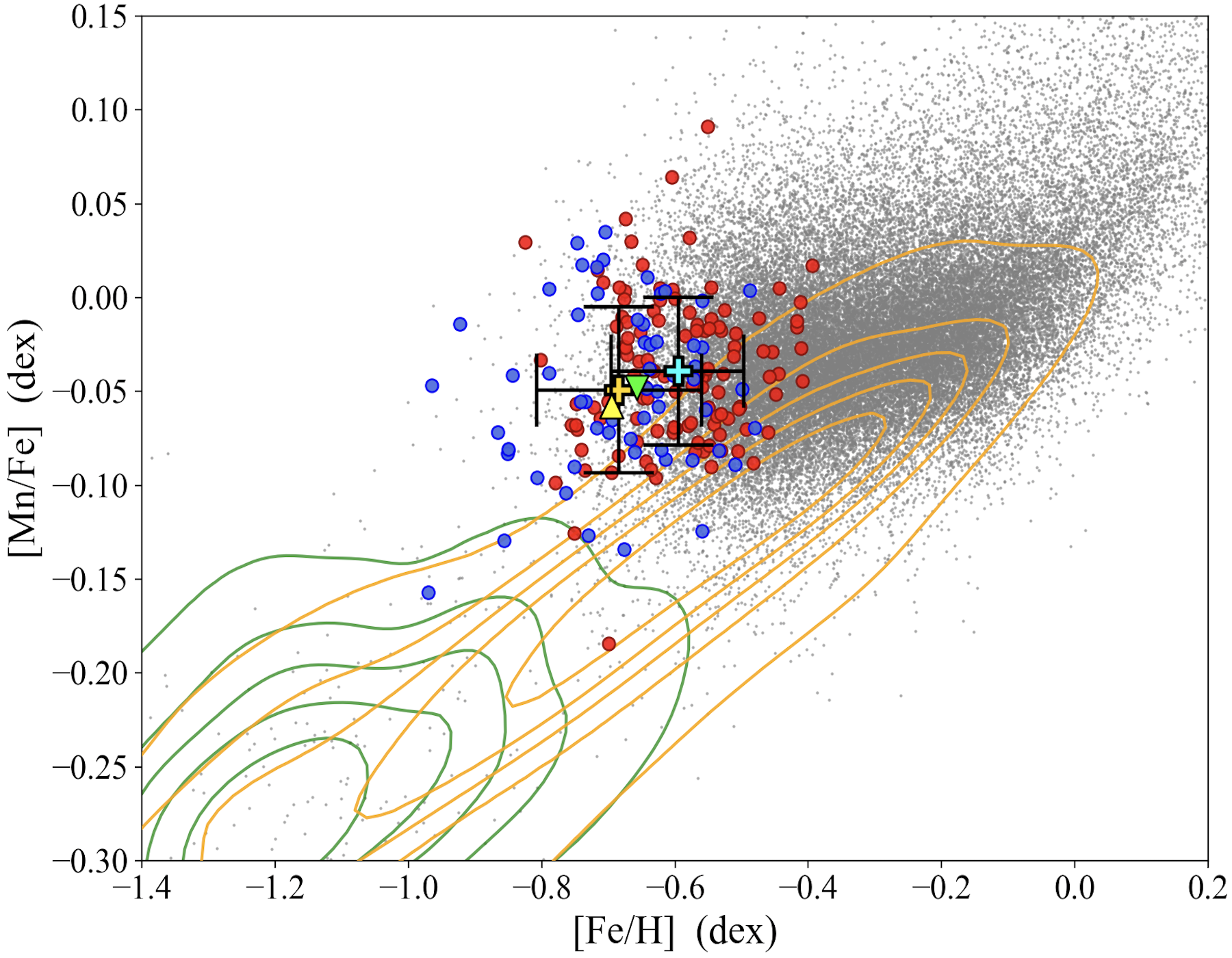}
	\hspace{0.34cm}
	\includegraphics[scale=0.458]{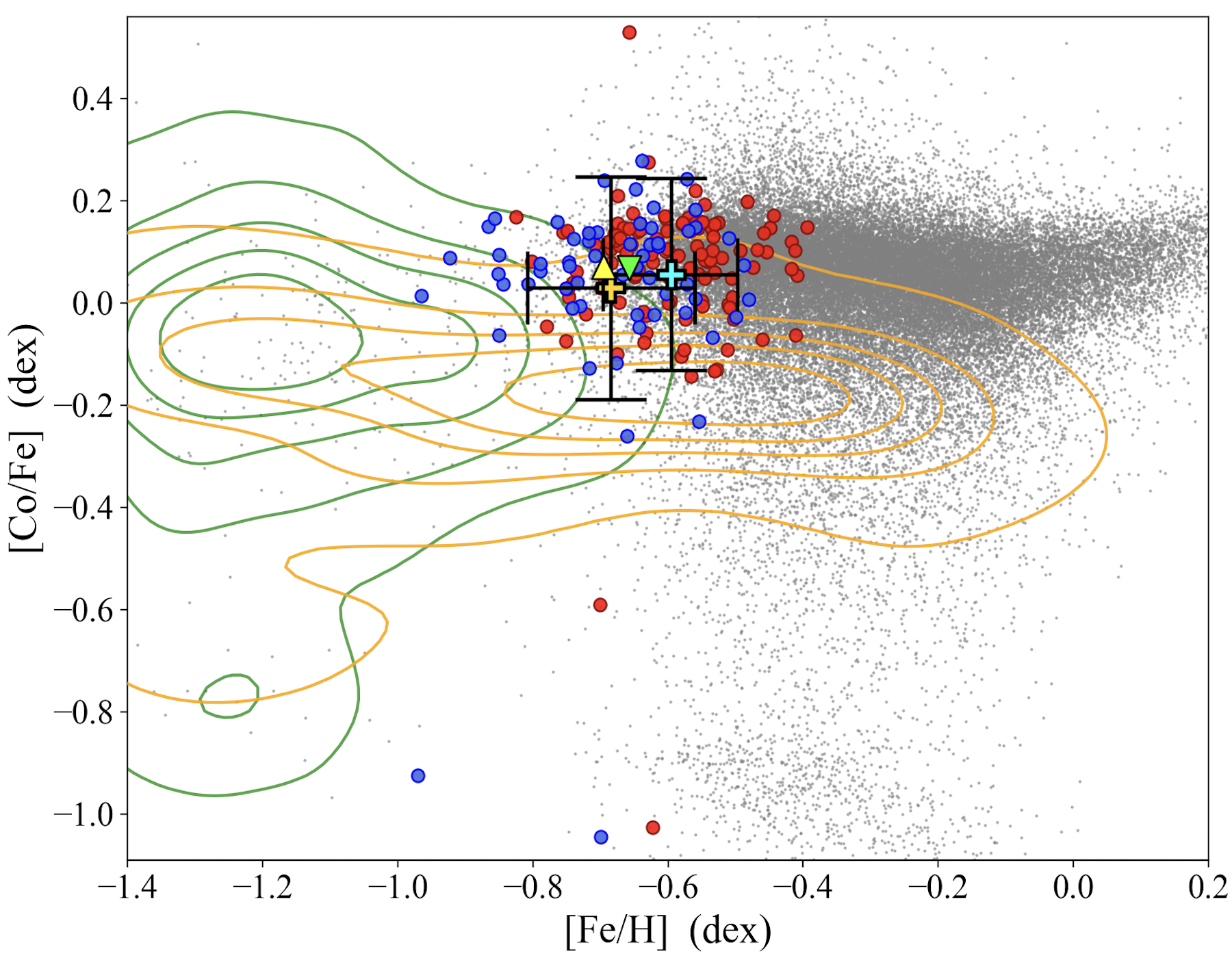}
	\caption{Abundance extrapolation results of other elements. All marks are same as \textbf{Figure \ref{fig5}}. }
	\label{figa1}
\end{figure*}

$\hspace*{0.2cm}$In order to fit the [X/Y]-$R_{GC}$-$Z_{GC}$ relation for the outer thin disk, we spatially divide the thin disk sample into a grid of $R_{GC}$ and $|Z_{GC}|$: $10<R_{GC}<18$ kpc, with step size of 1 kpc; $0<|Z_{GC}|<3$ kpc, with step size of 0.5 kpc. For each 1 kpc $\times$ 0.5 kpc cell, we calculate the mean abundance of each element, mean $R_{GC}$ and $|Z_{GC}|$. Interestingly, \cite{2022ApJ...928...23E} showed that the chemical abundance trends of MW disk deviate from linear fitting in 10 $<R_{GC}<$ 18 kpc. Therefore, we explore several combination of mathematical functions besides the traditional linear fitting function: [X/Y] = a$\times$$R_{GC}$ + b$\times$$|Z_{GC}|$ + c. In the end, we find that the best fitting function is: [X/Y] = a$\times\sqrt{R_{GC}}$ + b $\times \sqrt[4]{|Z_{GC}|}$ + c , where $a, b, c$ are coefficients determined by APOGEE data (see Table \ref{tab:1}). Let us take [Mg/Fe] as an example here (other elements show similar results). Figure \ref{fignew} shows that our proposed function shows smaller residuals than the linear function, especially at larger $R_{GC}$ and $|Z_{GC}|$, which means that our proposed function is better at predicting the chemical abundances of MW disk around the locations of these two substructures. To statistically quantify the residuals, we use two information criteria, AIC (Akaike information criterion) (and Chi-square test) gives -6.61676 and -6.46612 (0.015642 and 0.017922) for our proposed function and the linear function, respectively. In both tests, our proposed function performs better than the linear function, thus we will use our proposed function for the following discussion. 

$\hspace*{0.2cm}$ After that, our fitting functions are extrapolated to the $R_{GC}$ and $Z_{GC}$ of each ACS or MNC star in order to obtain its predicted abundances, then we calculate the two samples' mean predicted abundances of each element to compare with their mean observed abundances. If ACS and MNC are chemically associated with the thin disk, the observed abundances should agree with the predicted values within the measurement uncertainties. The observations generally confirm this idea: The ACS and MNC stars are generally located near the very low end of thin disk metallicity distribution, and their element abundances do not strongly deviate from other thin disk stars around their metallicity range. For all the elements (including Fe) we investigate here except Mg, the differences between the predicted mean abundances and corresponding observed mean values are smaller than 1$\sigma$ (\textbf{Figure \ref{fig5}} for representative elements, also see \textbf{Figure \ref{figa1}}). As for Mg, this difference is larger than 1$\sigma$ for ACS, but still within 1.4$\sigma$ and no more than 0.03 dex (see \textbf{Section \ref{sec:4}} for further discussion). The mean [X/Fe] and metallicities are similar between ACS and MNC, proving that these two substructures are chemically similar to each other. On the other hand, ACS and MNC stars show little overlap with low-mass dwarf galaxies (Sgr and GSE, green and orange contours) in most chemical abundance spaces, which clearly excludes the possibility that these two substructures are originated from dwarf galaxies. 
\begin{itemize}
	\item C, N, O could be significantly affected by the H-burning CNO cycles, so they may be altered during the stellar evolution. In the ASPCAP, O abundances are determined from OH lines, C from CO lines, and N from CN lines. Therefore, N abundances suffer from the largest uncertainties and this may explain the larger dispersion of [N/Fe] compared to other light elements. 
	\item $\alpha$ elements are mainly generated in massive stars through Type II supernovae (SNe II). Mg and Si abundances are relatively well-determined, given their strong feature lines in the APOGEE spectra, and thus their abundances show small scatters for ACS and MNC member stars. On the contrary, Ca lines are generally weaker, causing larger scatter in their abundances. 
	\item Odd-Z elements, including Al and K that we investigate in this work, are mostly generated through the proton-capture process during SNe II. Al is one of the most reliable elements from the APOGEE spectra, thanks to its strong feature lines. On the other hand, K is less well-constrained with four MNC stars having particularly low K abundances. 
	\item Iron-peak elements are in fact a heterogeneous combination of both SNe Ia and core-collapse SNe II, where the former contributes most significantly. Cr and Co are good tracers of Fe, as their [X/Fe] are almost flat as a function of metallicity for thin disk stars, although it would also cause severe overlap with GSE dwarf galaxy. 
	The similar trends between [Ni/Fe] and [$\alpha$/Fe] indicates low contribution of Ni from Type Ia supernovae (SNe Ia) \citep{2011A&A...530A..15N}, while the increasing trend of [Mn/Fe] as a function of [Fe/H] implies a metallicity-dependent Mn yield from SNe Ia. Ni and Mn are relatively well-constrained in this metallicity range with reasonable scatters. 
\end{itemize} 

$\hspace*{0.3cm}$In conclusion, our results suggest that ACS and MNC chemically matches the extrapolation of outer thin disk stars, based on distances and heights. The similar chemical pattern suggests that ACS and MNC have similar star formation history as the MW outer thin disk. This provides the strongest evidence of their association with the MW thin disk, and unambiguously excludes their possible link with dwarf galaxies.

\section{Discussion}
\label{sec:4}

\subsection{Chemo-dynamical Signatures}

$\hspace*{0.2cm}$The combined messages from dynamics and chemistry are important signatures that reveal the formation details of our Galactic disk. Located at distances far above the Galactic plane, whether ACS and MNC belong to the Galactic thin disk has been a puzzle for a long time. Their dynamical properties, Mg and Fe were found to be consistent with the thin disk in the works of \cite{2020MNRAS.492L..61L} and \cite{2021ApJ...910...46L}. To further investigate if ACS and MNC share the same star formation history and thus the same chemical patterns as the Galactic thin disk, we definitely need to explore more elements. Toward this, \cite{2010ApJ...720L...5C} measured the Ti, Y and La abundances of MNC, and argued their chemical trajectories agree with the trends outlined by Galactic stars. Our work gives a more complete picture to this investigation: we confirmed that ACS and MNC are dynamically consistent with the Galactic thin disk. More importantly, we found that these substructures also share the same chemical patterns as the thin disk, including 12 spices of light elements, alpha elements, odd-Z elements, and iron-peak elements. 

$\hspace*{0.2cm}$ACS and MNC stars are located in the low-metallicity, high-$\alpha$ end of the MW thin disk sequence, indicating that these two substructures were formed prior to most thin disk stars. Interestingly, there is a possible flattening trend of outer thin disk around ACS and MNC in the [$\alpha$/Fe] \textit{vs} [Fe/H] plane, as the predicted mean [$\alpha$/Fe] are higher than observed mean values\footnote{This flattening trend is even clearer when linear fitting function is used to predict chemical abundances.}. The most significant case is [Mg/Fe], where their difference is larger than 1$\sigma$ for ACS. However, we again acknowledge that this phenomenon is weak, as the excess is no larger than 1$\sigma$ for most cases. This possible feature requires further investigated when more off-plane stars at larger $Z_{GC}$ and $R_{GC}$ are observed. 

\begin{figure}[htbp]
	\raggedright
	\includegraphics[scale=0.458]{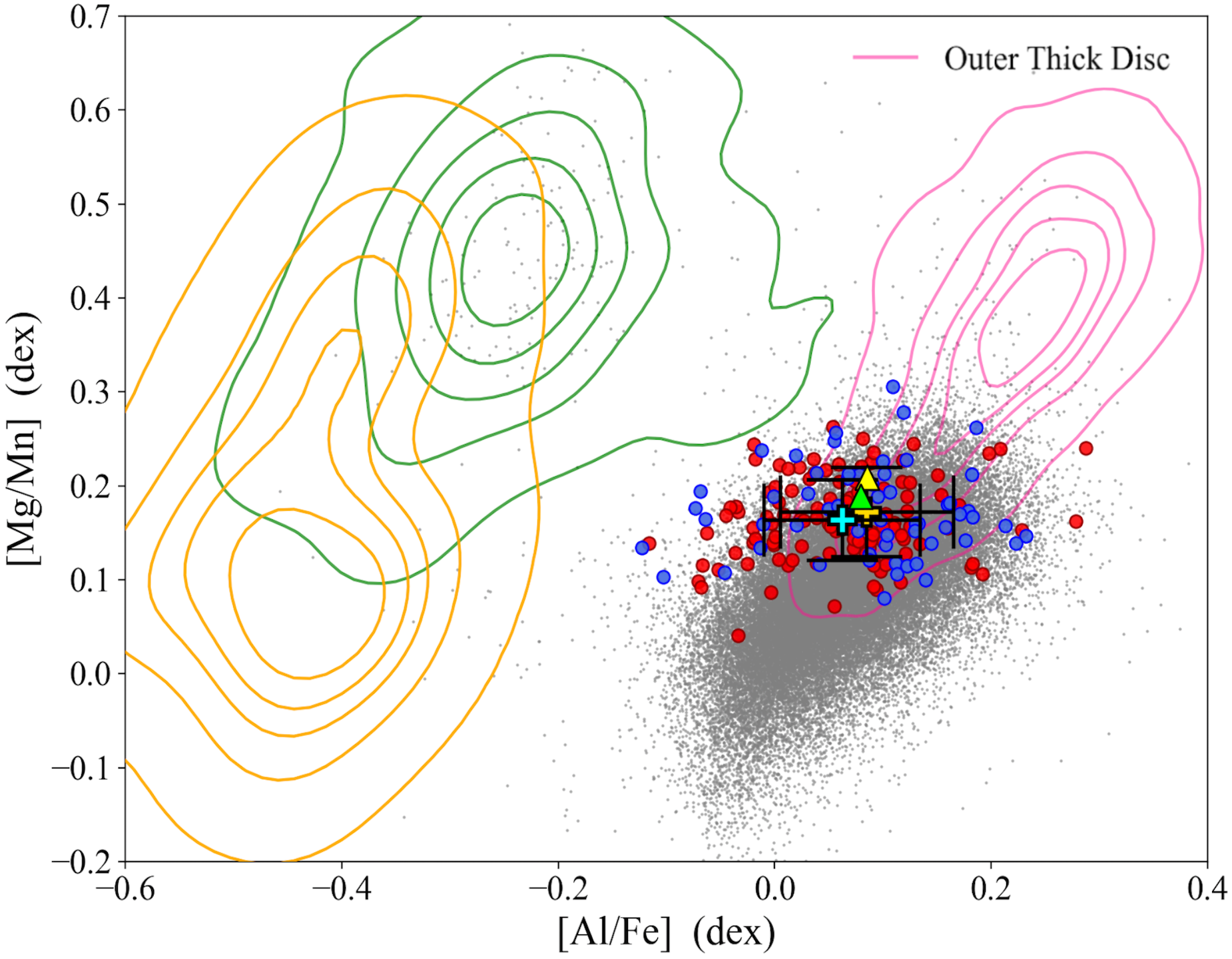}
	\caption{The extrapolation results on the plane of [Mg/Mn] and [Al/Fe]. The magenta isodensity contours are associated with stars divided into the thick disk part above, and other marks are same as \textbf{Figure \ref{fig5}}. }
	\label{fig6}
\end{figure} 

$\hspace*{0.2cm}$Besides the traditional [X/Fe] \textit{vs} [Fe/H] plane, the [Mg/Mn] \textit{vs} [Al/Fe] plane has also been proposed recently \citep{2020MNRAS.493.5195D}, where MW disk, MW halo, dwarf galaxies could be clearly separated. ACS and MNC stars are located near the high [Mg/Mn] and high [Al/Fe] end of the thin disk star distribution, partially overlapping with other thick disk stars (see \textbf{Figure \ref{fig6}}). According to the simulated MW evolution trajectories of \citet{Horta2021}, this suggests the stars of these two substructures predate most thin disk stars, in accordance with what we found above. Moreover, these two substructures are clearly separated from dwarf galaxies in the [Mg/Mn] \textit{vs} [Al/Fe] plane, confirming again their lack of connection. 

\subsection{Formation Scenarios}

$\hspace*{0.2cm}$As shown above, ACS and MNC are dynamically and chemically similar to the MW thin disk. However, how could these substructures be found so far away from the Galactic midplane ($Z_{GC}$ can reach 8 kpc)? One of the most popular scenarios is that they rose to current heights several hundred Million years ago when the MW disk was vertically flared due to the last passage of Sgr  \citep[e.g.,][]{2018MNRAS.481..286L,2019MNRAS.485.3134L,Xuyan2020}. This scenario simultaneously explains the observed phase spiral structure found in Z \textit{vs} $V_Z$ space \citep{2018Natur.561..360A}, further increasing its credibility. Another hypothesis was mentioned by \cite{2022MNRAS.513.4130L} that the geometrically thick disk was formed in two episodes. In the first episode, the inner part of the thick disk was born as the classical thick disk with higher $\alpha$ abundances. Later in the second episode, the outer part was formed along with the geometrically thin disk but earlier, and this part has similar chemical evolution trajectories to the classical thin disk with lower $\alpha$ abundances, but it's much further from the MW midplane than the thin disk. These two scenarios both assume that the off-plane substructures are chemically similar to the MW outer thin disk, which is unequivocally shown in our work.

$\hspace*{0.2cm}$The next major question is when were the stars of these substructures born? Based on spectro-phometric ages derived using Bayesian probability density functions \citep{2018MNRAS.481.4093S}, \cite{2020MNRAS.492L..61L} found that ACS is predominantly populated by stars with ages around 10 Gyr, whereas MNC has a larger proportion of younger stars (6-10 Gyr). Given that most of these stars were born before the first passage of Sgr (5-7 Gyr ago,  see \textbf{Figure 2} of \citealt{2020NatAs...4..965R}), these authors suggested ACS was kicked up shortly after the first Sgr passage, while MNC was gradually built up through a succession of encounters. 

\begin{figure}[htbp]
	\centering
	\includegraphics[scale = 0.55]{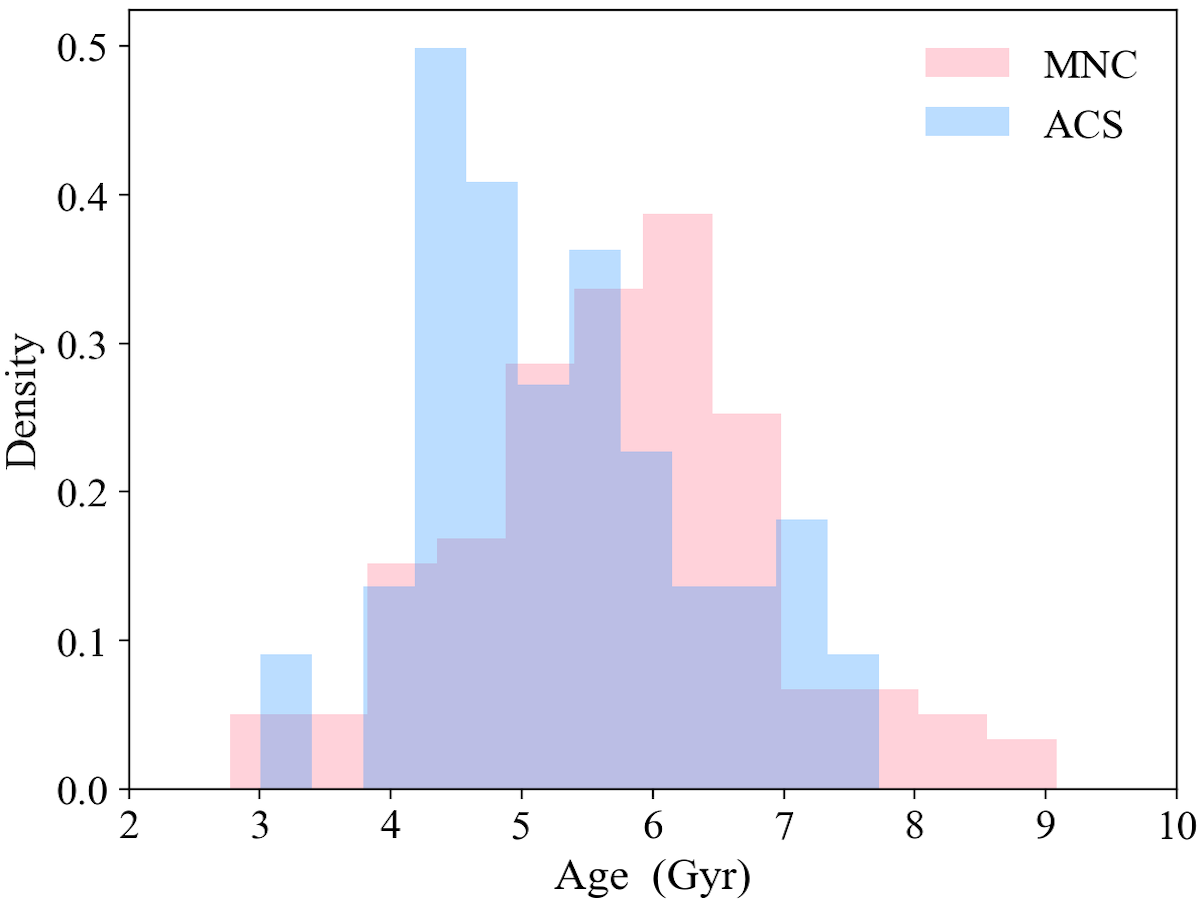}
	\caption{Age distributions (from AstroNN) exhibited in histograms, lightblue for ACS and pink for MNC stars. }
	\label{fig7}
\end{figure}

$\hspace*{0.2cm}$To examine the age distributions of our sample stars, we retrieved ages from AstroNN, which are estimated using a Bayesian neural network model trained on asteroseismic ages \citep[see][]{2019MNRAS.489..176M}. The typical age model error of our sample stars is around 1 Gyr. The age distributions (\textbf{Figure \ref{fig7}}) show that the majority of both ACS and MNC candidates are between 4-7 Gyr (5.324 $\pm$ 1.062 Gyr for ACS stars, 5.752 $\pm$ 1.178 Gyr for MNC stars). The age distributions from AstroNN are significantly different than those from \citet{2018MNRAS.481.4093S}. The latter one predicted larger distances, and thus larger Galactic heights compared to the former one. Given that the Galaxy prior employed in \citet{2018MNRAS.481.4093S} does not include disk undulation and flares, we speculate the associated distances and ages might be overestimated. Interestingly, the stellar ages (from AstroNN) of these two substructures coincide with the time when the first Sgr encounter happened (5-7 Gyr ago). \citet{2020NatAs...4..965R} revealed three episodes (5.7, 1.9, 1 Gyr) of enhanced star formation for stars within 2 kpc around the Sun, and suggested that they may be triggered by Sgr pericentric passages.  Our newly estimated ages indicate that stars in ACS and MNC were formed during the first Sgr pericentric passage, but whether they were kicked up to their current heights by the first or later Sgr passages requires careful dynamical modelling, which is beyond the scope of this work. 

\section{Summary}
\label{sec:5}

$\hspace*{0.2cm}$As more and more photometric, astrometric, and spectroscopic surveys provide measurements for faint stars, the astronomical community start to appreciate the complex substructures of the outer Galactic disk: flare, warp, undulation and etc. Whether the off-plane ACS and MNC substructures are part of the Galatic thin disk has been debated for long. In this work, we selected reliable candidates of ACS and MNC from the APOGEE survey, where accurate radial velocities and chemical abundances are available. The ACS and MNC stars show nearly circular orbits, with consistent orbital energy and vertical angular momentum as MW outer disk stars. By extrapolating chemical abundance trends in the outer thin disk region ($10 < R_{GC} < 18$ kpc, $0 < Z_{GC} < 3$ kpc), we found that ACS and MNC stars show consistent chemical abundances as the extrapolating values for 12 elements. In other words, ACS, MNC and MW outer thin disk are chemically similar to each other. Moreover, ACS and MNC are chemically separable to dwarf galaxies, like GSE and Sgr, disproving their association. Furthermore, we found that the ages of ACS and MNC stars are mostly between 4-7 Gyr, which coincides with the time of the first Sgr encounter. 

\begin{acknowledgements}
	$\hspace*{0.4cm}$We thank Chao Liu for insightful discussions. Y.Qiao, B.Tang and C.Xu gratefully acknowledge support from the Natural Science Foundation of Guangdong Province under grant No. 2022A1515010732 and the National Natural Science Foundation of China under grant No. 12233013. J.Lian acknowledges support from the Applied Basic Research Foundation of Yunnan Province under grant No. 202105AE160021 and No. 202005AB160002. J.Li acknowledges support from the National Natural Science Foundation of China under grant No. 12273027. 
\end{acknowledgements}

\software{TOPCAT \citep{taylor2005}, matplotlib \citep{Hunter2007}, numpy \citep{numpy2011,numpy2020}, math, glob, pandas \citep{pandas2010}, astropy \citep{astropy2013,astropy2018}, scipy \citep{scipy2020}, seaborn \citep{seaborn2021}, galpy \citep{2015ApJS..216...29B}, galstreams \citep{2023MNRAS.520.5225M}, StarHorse \citep{starhorse2018}. }

\bibliography{sample}{}
\bibliographystyle{aasjournal}

\end{document}